# Terahertz generation in Czochralski–grown periodically poled $Mg:Y:LiNbO_3$ via optical rectification

G. H. Ma[1*], G. Kh. Kitaeva[2**], I. I. Naumova[2], S. H. Tang[1]

[1] Department of Physics, National University of Singapore, 119260, Singapore

[2] Faculty of Physics, Moscow State University, 119899, Moscow, Russia

Abstract

Using a canonical pump-probe experimental technique, we studied the terahertz (THz) waves generation and detection via optical rectification and mixing in Czochralski-grown periodically poled $Mg:Y:LiNbO_3$ (PPLN) crystals. THz waves with frequencies at 1.37 THz and 0.68 THz as well as 1.8 THz were obtained for PPLN with nonlinear grating periods of 30 and 60 μm, respectively. A general theoretical model was developed by considering the dispersion and damping of low frequency phonon-polariton mode. Our results show that THz waves are generated in forward and backward directions via pumping pulse rectification. The generated THz waves depend on the spectral shape of the laser pulses, quasi-phase mismatches and dispersion characteristics of a crystal.

[*] Email address: phymagh@nus.edu.sg
[**] Email address: kit@qopt.phys.msu.su

## I. INTRODUCTION

Generation and control of terahertz (THz) frequencies radiation remains the area of intensive research during the last decade. This range of electromagnetic spectrum is very attractive due to a variety of promising applications in spectroscopy, outerspace communication, biomedical imaging and tomography.[1, 2] Generally, there are two methods for THz wave generation. The first method employs ulatarshort laser pulse to illuminate a photoconducting semiconductor to induce a transient electrical current which results in THz emission[3]. The other method is via optical rectification in nonlinear dielectric crystals pumped by ultrashort laser pulses [4]. The later method can be used to generate broad-band, single-cycle THz waves under phase matching condition,[5, 6] as well as to generate narrow-band, multicycle THz waves under quasi-phase matching condition.[7, 8]

As a femtosecond pulse illuminates a nonlinear crystal such as $LiNbO_3$, ZnTe etc., a THz nonlinear polarization is generated via optical rectification. The ultrashort-pulse induced nonlinear polarization follows the intensity profile I(t) of the pulse. Ideally, for single-cycle THz generation, phase matching condition must be met. The group velocity of the optical and THz pulses should be the same in the crystal so that THz field emitted by each dipole of the crystal adds coherently in forward direction. If there is a group-velocity mismatch, the optical and THz pulses will walk off each other as they propagate through the crystal. As a result, the only nonzero contribution of $P^{(2)}(t)$ to the far-field radiation comes from the regions within a coherent length near the front and back surfaces of the crystal. If the nonlinear polarization of the crystal is periodically reverse to compensate the walk-off effect due to the group-velocity mismatching, a narrow-band or complex multicycle THz wave form can be generated. In order to describe precisely the THz wave generation and propagation in a ferroelectric crystal such as $LiNbO_3$ as well as Mg-doped $LiNbO_3$, the dispersion and damping properties of the crystal has to be included because of the strong phonon polariton coupling between phonon and infrared light. [9, 10]

$LiNbO_3$ crystals, being among the most effective nonlinear crystals for frequency conversion in the visible and IR range, have already demonstrated their excellent properties for THz generation. A tunable terahertz parametric oscillators (TPO) under near-infrared (IR) nanosecond-pulse pumping was proposed and demonstrated based on $LiNbO_3$ crystal.[11, 12] It was shown, that the more than 10 times enhancement of THz-wave output is achieved by using double-pass pumping geometry and Mg-doped $LiNbO_3$ crystals.[13] In 2000, Lee et al demonstrated that a narrow-band THz radiation was obtained via optical rectification in periodically poled $LiNbO_3$ (PPLN) crystals, where the domain length is matched to the walk-off length between the optical and THz pulses.[7] The advantages of PPLN for the generation of THz are its abilities to simplify the optical scheme, to enlarge the tunability of THz spectrum and to obtain narrower bandwidths. Recently, the new tuning scheme was realized to obtain narrow band THz radiation in disk-shaped PPLN,[14] where domain grating was induced in thin crystal layers fabricated by postgrowth poling technique.

In the present work we study the THz-generation and detection of the Mg-doped PPLN crystals. The Mg-doping is known to protect crystals from optical damage[15] and promises to increase the THz-output as in TPO experiments.[13] Our crystals contained regular domain grating are produced directly under Czochralski growth procedure.[16, 17]

Periodically poled samples of larger sizes with plane domain walls can be obtained by this method and are reported to be stable upon possible thermal and radiation influence.[16-18] It is a well-known fact that periodically poled crystals made by post-growth techniques have limited sizes along the optical Z-axis direction.[19] Being no more than 0.5-1 mm, these sizes are almost comparable with THz wavelengths. On the contrary, the as-grown periodically poled crystals are free of this disadvantage. The sizes of the elements in all directions can be 5-10 cm and more. This promises the possibility of more efficient generation and detection of THz-range waves via optical rectification and other nonlinear processes. For the first testing of the THz-generation capabilities, we used the canonical femtosecond pump-probe technique. This technique provides an overall observation of ultrafast dynamics of different effects, both coherent and incoherent, which can take place under powerful short-pulse pumping of a new material.[20] Instead of far-field electro-optic sampling method which can only detect the forward propagating THz wave, this method is an all-optical method, which can detect THz wave propagation in both forward and backward direction.

This paper reports the theoretical and experimental studies of THz wave generation, propagation and detection in Mg-doped PPLN crystal by considering phonon polariton dispersion and damping. In Sec. II we will develop the general expression of THZ wave generation, frequency mixing between generated THz wave and probe beam, and THz detection using lock-in technique. In Sec. III, we describe the fabrication of samples, pump-probe experimental setup, and results of experiments. Discussions are given in the Sec. IV and conclusions are presented in Sec. V.

## II. THEORITICAL TREATMENT

In this section, we present the theoretical consideration for THz generation and detection in a canonical pump-probe configuration with lock-in technique. To simulate the experimental data, the following three steps have to be considered. (A) pump-beam induce THz generation via optical rectification; (B) frequency mixing between generated THz wave and probe in nonlinear crystal, and (C) THz wave detection via lock-in technique.

### **A.** THz wave generation in PPLN crystal under power pumping via optical rectification.

In one-dimensional approximation, the amplitude $E_{TH}(\Omega, x)$ of each spectral component $E_{TH}(\Omega, x)e^{-i\Omega t}$ at the THz frequency $\Omega/2\pi$ is dependent on the coordinate x along the crystal X-axis according to the general wave equation:

$$\frac{\partial^2 E_{TH}(\Omega, x)}{\partial x^2} + \varepsilon(\Omega)\frac{\Omega^2}{c^2} E_{TH}(\Omega, x) = -\frac{4\pi\Omega^2}{c^2} P_I^{(2)}(\Omega, x) \tag{1}$$

where $\varepsilon(\Omega)$ is the dielectric function of the crystal. The one-dimensional approximation is valid for small angles ($\theta$) of pump and probe incidence to the crystal. When $\theta \approx 0$ the nonlinear polarization in this step can be expressed as

$$P_I^{(2)}(\Omega, x) = \frac{\chi^{(2)}(x)C(\Omega)}{2\pi} \exp(i\frac{\Omega}{v_{gr}}x) \tag{2}$$

where C(Ω) is the Fourier transform of the optical pump pulse $\left|E_{pump}(t)\right|^2$.[21, 22] In Eq. (2) we neglect the spectral dispersions of the second-order susceptibility $\chi^{(2)}$ and of the group velocity $v_{gr}$ for the pump waves, but take into account the inhomogeneous distribution of $\chi^{(2)}(x)$ along X-axis. In the periodical structure of PPLN, $\chi^{(2)}(x)$ takes the form of the Fourier series

$$\chi^{(2)}(x) = \sum_{m=-\infty}^{\infty} \chi_m \exp(imqx) \tag{3}$$

where m=1, 2, 3…, q=2π/d is the unit of reciprocal vector and d is the period. The Fourier transform of $\chi^{(2)}$ is,[23]

$$\chi_m = \frac{\bar{\chi}(-1)^{mn}}{\pi m}\left\{(-1)^m \sin m\pi\rho + i\left[1-(-1)^m \cos m\pi\rho\right]\right\} \tag{4}$$

where $\rho \equiv (l_+ - l_-)/(l_+ - l_-) = 2l_+/(l_+ + l_-) - 1$ is a relative difference between the lengths of positive ($l_+$) and negative ($l_-$) domains, n=L/d is a number of periods in a crystal and $\bar{\chi}$ is the effective non-linear susceptibility of each domain. In our experiments the samples are designed with $\bar{\chi} = \chi_{ZZZ}^{(2)}$ and $\rho = 0$ (duty cycle is 1). As the first order harmonics are appreciably larger than the higher order ones, a good approximation can be attained by taking into account only the first harmonics $\chi_1$ and $\chi_{-1}$. Substituting (4) into (3) and considering m=1 only, $\chi^{(2)}$ (x) can be obtained in the form:

$$\chi^{(2)}(x) = \frac{2i\chi_{zzz}(-1)^n}{\pi}\left[\exp\left(i\frac{2\pi}{d}x\right) - \exp\left(-i\frac{2\pi}{d}x\right)\right] \tag{5}$$

In general, a resolution of the wave equation (2) yields two opposite THz waves, $E_{TH}(\Omega,x)\exp(-i\Omega t) = E_+(x)\exp(ikx - i\Omega t) + E_-(x)\exp(-ikx - i\Omega t)$. The amplitudes of forwards $E_+(x)$ and backwards $E_-(x)$ propagation THz waves are calculated as,[24]

$$E_+(x) = \frac{2\pi i\Omega^2}{kc^2}\int_0^x P^{(2)}(\Omega,x`)e^{-ikx`}dx`,$$

$$E_-(x) = \frac{2\pi i\Omega^2}{kc^2}\int_x^L P^{(2)}(\Omega,x`)e^{ikx`}dx`,$$

In the absorption region, the value of the wave vector $k = \frac{\Omega}{c}\sqrt{\varepsilon(\Omega)}$ is complex, $k = \Omega n(\Omega)/c + i\alpha(\Omega)/2 = k`+ik``$. The real part $k'$ is dependent on the dispersion of refractive index n(Ω), and the imaginary part $k''$ is related to absorption coefficient α(Ω). Substituting (4) into (2) and solving the wave equation (1), we obtain for the case of a periodically poled crystal:

$$E_{THz}(\Omega,x) = -\frac{2\chi_{zzz}(-1)^n\Omega^2 C(\Omega)}{\pi kc^2}\times$$
$$\begin{Bmatrix} x\exp(ik`x - k``x/2)\lfloor\exp(i\Delta_1 x/2)f(\Delta_1,x) - \exp(i\Delta_2 x/2)f(\Delta_2,x)\rfloor - \\ (L-x)\exp[-ik`x - k``(L-x)/2][\exp(-i\Delta_3(L+x)/2)f(\Delta_3,L-x) - \exp(-i\Delta_4(L+x)/2)f(\Delta_{4,}L-x)] \end{Bmatrix} \quad (6)$$

Here $f(\Delta_i,x)$ are form-factor functions with i=1, 2, 3, 4, which are determined as,

$$f(\Delta,l) = \frac{e^{(i\Delta+k``)l/2} - e^{-(i\Delta+k``)l/2}}{(i\Delta+k``)l}.$$

The values of $|f(\Delta_i,x)|$ reach maxima in phase matching conditions,

$$\Delta_1 = \frac{\Omega}{v_{gr}} + \frac{2\pi}{d} - k` = 0 \tag{7a}$$

$$\Delta_2 = \frac{\Omega}{v_{gr}} - \frac{2\pi}{d} - k` = 0 \tag{7b}$$

$$\Delta_3 = -\frac{\Omega}{v_{gr}} + \frac{2\pi}{d} - k` = 0 \tag{7c}$$

or

$$\Delta_4 = -\frac{\Omega}{v_{gr}} - \frac{2\pi}{d} - k` = 0 \tag{7d}$$

The resonance frequencies $\Omega_i$ are determined by equations (7a)-(7d) through $\Delta_i(\Omega_i)=0$. The widths and maximum values of form-factors are dependent on the absorption coefficient at the resonance THz frequency $\alpha(\Omega) = 2k``$. If the crystal is transparent at the THz frequency, i.e. k``=0, the form-factor function is reduced to sinc function, i. e. $f(\Delta,0,l) = \sin c(\Delta l/2)$.

In general case, the Eqs. (7a), (7b), and (7c) can be satisfied in a periodically poled crystal of a proper period, while Eq. (7d) can never be satisfied. In pumping a periodically poled crystal, it is possible to generate one backward and two forward THz waves. The spectral width and the total intensity of each generated beam are dependent on the value of absorption $\alpha(\Omega)$ and the spectral component of the pumping pulse $C(\Omega)$. In the derivation of Eq. (6), we have assumed that the incident pump beam is a plane wave in a crystal which has total length L along its X-axis. When the pump beam is focused critically, the main part of a terahertz field is generated in a focal region and L can be taken close to a confocal parameter.

B. Parametric interaction between the THz and probe beams

Since $LiNbO_3$ crystals have extremely high second-order susceptibility, we will neglect $\chi^{(3)}$-based contributions and take into account only the processes due to $\chi^{(2)}$. The transmittance of probe beam will be modulated after considering the interaction between the generated THz and probe beam. The slowly varying amplitude $E_{pr}(\omega, x)$ of each spectral component of the probe wave $E_{pr}(\omega,x)\exp\left(i(k_{pr}(\omega)x-\omega t)\right)$ at frequency $\omega$ yields,

$$\frac{\partial E_{pr}(\omega,x)}{\partial x}=\frac{2\pi i\omega^2}{k_{pr}c^2}P_{II}^{(2)}(\omega,x)\exp(-ik_{pr}x) \tag{8}$$

where the non-linear polarization in this parametric interaction is

$$P_{II}^{(2)}(\omega,x)=\frac{i\chi_{zzz}(-1)^n}{\pi^2}\left(\exp\left(i\frac{2\pi}{d}x\right)-\exp\left(-i\frac{2\pi}{d}x\right)\right)$$
$$\times\left[\int_0^\infty E_{pr}(\omega+\Omega,x)\exp(ik_{pr}(\omega+\Omega)x)E_{TH}^*(\Omega,x)+E_{pr}(\omega-\Omega)\exp(ik_{pr}(\omega-\Omega)x)E_{TH}(\Omega,x)\right]d\Omega \tag{9}$$

Equation (9) shows that two major contributions are accountable for the mixing processes: the first one arises from the difference frequency mixing, and the second one is as the result of sum frequency mixing between the spectral components of the probe beam and the THz waves. Integration is carried out along the whole spectrum of the generated THz waves in step A. Again, the spatial dependency of the second-order susceptibility ($\chi^{(2)}$) was taken as in form (4). For simplification, the wave vectors of $k_{pr}(\omega+\Omega)$ and $k_{pr}(\omega-\Omega)$ were approximately expressed as $k_{pr}(\omega)+\Omega/v_{gr}$ and $k_{pr}(\omega)-\Omega/v_{gr}$, respectively.

The amplitudes of the probe waves in Eq. (8) can be regarded as $E_{pr}(\omega,x)=E_{pr}(\omega)+\Delta E_{pr}(\omega,x)$, where $E_{pr}(\omega)$ is the same in all points of the crystal, and the non-linear additive $\Delta E_{pr}(\omega)$ due to the mixing with THz waves is small, i.e. $\left|\Delta E_{pr}(\omega,x)\right| << \left|E_{pr}(\omega)\right|$. Solving Eq. (8) in a first approximation, we neglected the small part $\Delta E_{pr}(\omega)$ in Eq. (9) for the nonlinear polarization. Finally, the $\Delta E_{pr}(\omega,L)$ at the output of a crystal can be obtained as $\Delta E_{pr}(\omega,L)=\int \Delta E_{pr}(\omega,\Omega,L)d\Omega$, where

$$\Delta E_{pr}(\omega,\Omega,L)=iLC(\Omega)\frac{4(\chi_{zzz})^2\Omega^2}{\pi^2c^3}\frac{\omega}{n(\omega)}\left[E_{pr}(\omega-\Omega)u(\Omega)+E_{pr}(\omega+\Omega)u^*(\Omega)\right], \tag{10}$$

with

$$u(\Omega)=\frac{1}{k}\times\left\{\begin{array}{l}\sum_{s=1}^{4}\frac{1}{\Delta_s-ik``}\\ +\frac{4\pi/d}{(\Delta_1-ik``)(\Delta_2-ik``)}\left[f(\Delta_1,k``,L)e^{-i(\Delta_1-ik``)L/2}-f(\Delta_2,k``,L)e^{-i(\Delta_2-ik``)L/2}\right]\\ +\frac{4\pi/d}{(\Delta_3-ik``)(\Delta_4-ik``)}\left[f(\Delta_3,k``,L)e^{-i(\Delta_3-ik``)L/2}-f(\Delta_4,k``,L)e^{-i(\Delta_4-ik``)L/2}\right]\end{array}\right\} \tag{11}$$

As carried out in step A in the derivation of Eq. (10), assumed plane-wave approximation and collinear propagation of pump and probe beams. Here, the parameter L denotes the interaction length between the probe and THz waves, which was taken to the same value as the THz wave generation in step A. If the pump and probe beams are non-collinear (θ≠0), the effective overlapping length of two beams will be deduced. In our experimental configuration, the angle between the pump and probe beams is less than 10 degrees. We can thus consider the two beams to be propagating in parallel direction (θ=0) for practical purpose. This process of the interaction between terahertz and probe waves can therefore be regarded as the detection of terahertz radiation. As different from the usual detection where a detector is space-separated from a source of radiation, here we have a detector built in an emitting crystal. Again, if a short focusing is used, the length of focusing regions should be taken into account.

C. Measurement of transmittance changes in the probe beam

In our experiment, we measure the pump-induced additive to the main probe signal, $\Delta T = \int dt\left[I(t,\tau_{del}) - I_0(t)\right]$. Integrating is made over the probe pulse duration. $I(t, \tau_{del})$ stands for the intensity of the probe wave in the crystal with pumping wave presence, which is proportional to $\left|\int_{-\infty}^{\infty} d\omega\left[E_{pr}(\omega) + \Delta E_{pr}(\omega, L)\right]e^{-i\omega t}\right|^2$, and $I_0(t) = \left|\int_{-\infty}^{\infty} d\omega E_{pr}(\omega)e^{-i\omega t}\right|^2$ stands for the intensity of the probe wave in the crystal without any previous pumping. By taking into account that $\Delta E_{pr}(\omega, L) << E_{pr}(\omega, L)$ for any delay time $\tau_{del}$ between pump and probe waves, we have

$$\Delta T = \int dt\left[\left|\int_{-\infty}^{\infty} d\omega\left[E_{pr}(\omega) + \Delta E_{pr}(\omega, L)\right]e^{-i\omega t}\right|^2 - \left|\int_{-\infty}^{\infty} d\omega E_{pr}(\omega)e^{-i\omega t}\right|^2\right]$$

$$\approx \int dt\left[\int_{-\infty}^{\infty} d\omega E_{pr}(\omega)e^{-i\omega t}\int_{-\infty}^{\infty} d\omega` \Delta E^*_{pr}(\omega`, L)e^{i\omega` t} + c.c.\right]$$

$$= \int dt\left[\int_{-\infty}^{\infty} d\omega \int_{-\infty}^{\infty} d\omega` \int d\Omega \Delta E^*_{pr}(\omega^*, \Omega, L)E_{pr}(\omega)e^{i(\omega` - \omega)t} + c.c.\right]$$

and taking into account the relation (10) for $\Delta E_{pr}(\omega, L, \Omega)$, we obtain,

$$\Delta T = L\frac{(\chi_{zzz})^2}{n_{opt}}\int_{-\infty}^{\infty} d\Omega E(\Omega)\Omega^2 u(\Omega)\int dt e^{-i\Omega t}\frac{\partial}{\partial t}\left|E_{pr}(t)\right|^2 \qquad (12)$$

Usually, the temporal shape of the both pump and probe pulses coincide, and their intensity profiles are proportional to $\left|E_{pump}(t)\right|^2 = I_{pump}F(t)$ as well as to $\left|E_{probe}(t)\right|^2 = I_{pr}F(t-\tau_{del})$, respectively. Here $\tau_{del}$ is the delay time between probe and pump beams, and the shape-function F(t) is connected with C(Ω) by the general relation $C(\Omega) = \frac{1}{2\pi}\int_{-\infty}^{\infty} dt I_{pump}F(t)e^{i\Omega t}$. Then Eq. (12) can be rewritten as:

$$\Delta T(\tau_{del}) = L\frac{I_{pr}}{I_{pump}}\frac{(\chi_{zzz})^2}{n_{opt}}\int_{-\infty}^{\infty} d\Omega\left|C(\Omega)\right|^2 \Omega^3 u(\Omega)\exp(-i\Omega\tau_{del}) \qquad (13)$$

Here $n_{opt}$ is the optical refractive index at laser wavelength (~800 nm), and $\Delta T(\tau_{del})$ describes the delay-time-dependence signal measured experimentally in arbitrary units. The spectral components of THz can be well resolved from the time-dependences transmittance changes of probe beam, $\Delta T(\tau_{del})$, by applying the reverse Fourier transform. In particular, according to Eq. (13), the amplitudes of spectral Fourier harmonics $\Delta T_\Omega$ should be dependent on THz frequencies Ω as

$$\Delta T_\Omega = [C(\Omega)]^2 \Omega^3 \left|u(\Omega)\right| \qquad (14)$$

Namely these amplitudes can be determined from the experimental time-dependence transmittance changes of probe beam, i.e. $\Delta T(\tau_{del})$ (Figs. 1b, 2b), the details of conclusions will be presented in the discussion section.

## III. EXPERIMENTS

Bulk periodically poled $Mg:Y:LiNbO_3$ single crystals were grown using the Czochralski method along X-axis from close to congruently melting composition Li/Nb

= 0.942 with 1 Wt % $Y_2O_3$ and 2 mol% MgO.[17, 18] Doping by Y was used to form periodically poled domain structure on the basis of rotation-induced growth striations in the bulk of $LiNbO_3$ single crystals. Codoping by Mg not only reduces the photorefractive damage, but also enhances the fixing of domain walls at the growth striations. Due to different ratio between pulling and rotation rates under the growth procedures, the crystals had different periods of nonlinear gratings of 30μm and 60μm. The samples were cut normally to X-axis and polished. Parallel to domain walls, the input and output surfaces of the samples were about 1cm×1cm. Sample 1 has domain period of 30 μm, containing more than 500 periods (with entire length along X-axis of 17mm), and sample 2 has domain period of 60μm, containing only 8 domain periods (with entire length along X-axis of 480 μm).

Pump-probe measurements were carried out on a canonical pump-probe setup.[25] The femtosecond laser pulse was generated from a Ti: Sapphire laser (Tsunami, Spectra-Physics) with pulse duration of about 200 fs, repetition of 82 MHz, and a centre wavelength at 800 nm. The laser beam was divided into pump (~90%) and probe (~10%) beams by a beamsplitter. The pump beam, with its polarization parallel to Z-axis of the crystal, was chopped at 1.7 kHz and it passed through an optical delayed line monitored by a computer-controlled step-motor. A quarter-wave plate (λ/4) and a polarizer were inserted into the probe beam path for adjusting the polarization of probe beam with respect to that of pump beam freely. Two beams, with separation of 12 mm, were focused on the same spot of a sample with a spot size of about 50 μm by a lens of f=50 mm. The peak intensity of the pump beam was ~ 3 $GW/cm^2$ at the sample position.

Effective THz-range modulation of the signal was obtained only when both the pump and the probe polarizations were taken parallel to Z-axis of samples. Fig. 1a and Fig.2a show the transient transmittance changes of probe beam for sample 1 and sample 2 , respectively. The wavelength of laser beam was tuned to 800 nm, and the angles between each beam and the crystal X-axis was about $\theta=5^{o}$, as shown in Fig. 1(c). In Figs. 1b and 2b we denote the results of Fourier transformation of these time-domain dependences. It is seen that the frequency of the most pronounced modulation is 1.37 THz in case of the sample 1 as well as 0.68 THz and 1.8 THz in case of sample 2.

IV. DISCUSSIONS

In order to simulate THz wave generation and propagation in the Mg-doped lithium niobate crystal, we have to consider the dispersion and damping of the THz wave vector because the crystal can not be treated as transparent in this particular electromagnetism wave region. In order to compare the experimental results with the theoretical predictions of Sec. II, we consider the dielectric function dispersion of Mg:Y:$LiNbO_3$ crystals in THz region in frames of a single polar phonon mode model, [26]

$$\varepsilon(\Omega) = \varepsilon_\infty + \frac{S_0\Omega_0^2}{\Omega_0^2 - \Omega^2 - i\Gamma\Omega} \qquad (15)$$

Here, $\Omega_0$, $S_0$ and $\Gamma$ are the transverse frequency, oscillator strength and dumping constant, respectively, of the lowest polar phonon, which determines substantially the crystal dispersion at a lower-polariton branch (its wavenumber is $\Omega_0/2\pi c$ = 256 $cm^{-1}$). As it was shown in ref. 27, to account for other low-frequency phonon excitations and to fit the experimental data on dispersion of refractive index and absorption coefficient in $LiNbO_3$ crystals, it is quite enough to consider the frequency dependence of a dumping constant $\Gamma$ in the following form,

$$\Gamma = \Gamma_0 + \frac{1}{i\Omega}\sum_j \frac{K_j}{\Omega_j^2 - \Omega^2 - i\Gamma_j\Omega} + \frac{1}{i\Omega}\frac{\Delta^2}{1-\Omega\tau} \tag{16}$$

Here, $\Omega_j$ and $\Gamma_j$ are the eigenfrequency and damping constant of the low-frequency excitation j, and $K_j$ represents the coupling strength.[27] For a Mg-doped congruent PPLN, an additional resonance frequency at $\Omega_1/2\pi c = 115\ cm^{-1}$ was accounted by taking $K_1/(2\pi c)^4 = 30.25\times10^6\ cm^{-4}$, $\Gamma_1/2\pi c = 45\ cm^{-1}$.[27] Since the Y-concentration in our crystals was very low, we neglected its possible influence on the crystal dispersion and considered the THz dispersion of both crystals (sample 1 and 2) in a same way. The other model parameters were $S_0=16.3$, $\Gamma_0/2\pi c = 17cm^{-1}\times(\Omega/\Omega_0)^3 + 9cm^{-1}\times(\Omega/\Omega_0)^4$, $\Delta/2\pi c = 26\ cm^{-1}$, $\tau=0.6$ ps.[26] The absorption coefficient $\alpha$ and refractive index n versus THz frequency $\Omega$, calculated according to Eq. (15) and (16) were plotted in Fig.3. It is seen that the absorption coefficient increase drastically when the frequency is larger than 2 THz. For example, absorption coefficient $\alpha$ is about 8 $cm^{-1}$ for $\Omega=1.37$, while the value of $\alpha$ is above 210 $cm^{-1}$ at $\Omega=3$ THz, which is why we unable to resolve the upper-frequency (about 3 THz) in sample 1. The values of real and imaginary parts of dielectric functions at each THz frequency $\Omega$ can be obtained based on Eq. (15) and (16), and then substituted to Eq. (11) to determine the values of $u(\Omega)$. The group velocity of periodically polled Mg:Y:$LiNbO_3$ crystals in the spectral range of pump and probe waves were determined using the data on the refractive index dispersion of our crystals, $v_{gr}=1.324\times10^{10}$ cm/s.[28]

Fig.4 (a) shows the frequency dependence of $\Delta T_\Omega = [C(\Omega)]^2 \Omega^3 |u(\Omega)|$, which is computed according to Eq. (14) for the sample 1 of Mg:Y: $LiNbO_3$ with grating periods 30 μm (L=1.7cm). The laser intensity profiles F(t) were taken as Gaussian distribution with pulse widths, 100 fs, 150 fs, and 200 fs, respectively. The spectral width of the factor $[C(\Omega)]^2$ is determined by the pulse duration, which limits the spectral range of the signal. The spectrum of $|u(\Omega)|$ in sample 1 consists of two different peaks. Position of the lower-frequency peak is determined by the quasi-phase matching condition Eq. (6c), and that of the upper-frequency peak is determined by Eq. (6a). Due to the lower crystal absorption at lower frequencies, peak of the first-type is appreciably sharper and its intensity is substantially (for more than an order of value) higher than that of the upper-frequency peak. While the factor $\Omega^3$ in Eq. (14) is seen to increase the relative amplitude of the higher frequency peak in $\Delta T_\Omega$, the factor $[C(\Omega)]^2$ is seen to suppress this peak heavily. As a result, in the case of 200 fs pulse duration the upper-frequency peak cannot be observed. This is confirmed very well by our experimental results (Fig. 4(a)). It is very interesting to mention that the most pronounced and highest peak observed in sample 1 corresponds to the THz wave, which propagates in inverse direction with respect to pump and probe beams. This THz wave can be detected directly at the input surface of the crystal. The experimental and predicted positions of this peak are in a very good agreement: 1.37 THz (experimental value) and 1.36THz (predicted value).

The power spectrum of sample 2 consists of two peaks with similar nature, as shown in Fig. 4(b). But due to a larger grating period, positions of the both peaks are shifted correspondingly to low frequency region, free of the limiting factor $[C(\Omega)]^2$ where absorption is smaller ($\alpha$<20 $cm^{-1}$). This agrees well with the observation that the upper frequency peak (corresponding to a forwards-propagating THz wave) is more pronounced and is observable. The frequencies of the experimentally observed and theoretically predicted positions of the upper-frequency peak are 1.8 THz and 1.75 THz, respectively. Again, the difference is lower than the predicted width of the peak (about 0.3 THz at a half-level). The calculated frequency of the backward-propagating THz wave peaks at 0.685 THz, which is very close to the experimental observation of 0.67 THz. It should be mentioned that, with pulse duration of 200 fs, the theoretical prediction that the intensity at lower-frequency should be more pronounced than that at upper-frequency which is not consistent with our experimental results. The reason is not clear at this moment. A possible reason, we surmise, is due to the poor periodicity of sample 2 which renders the frequency mixing between the probe and the 0.67 THz waves (step B, Sec. II for THz detection) to depart from the theoretical prediction.

## V. CONCLUSIONS

In conclusion, we have measured the THz response of Czochralski-grown periodically poled Mg:Y:$LiNbO_3$ crystals using a canonical pump-probe experimental technique. The well-pronounced THz modulation of the probe signal was observed at one preferred THz frequency for each sample: 1.37 THz for the sample with nonlinear grating period of 30 μm, and 0.67 THz and 1.8 THz for the sample with nonlinear grating period of period 60 μm. A general theoretical model was developed to explain the nature of signal formation. It was shown that pumping by femtosecond pulses leads to the generation of THz waves in a forward and backward directions, and that the probe beam can be regarded as internal detection of the both-type THz waves. The simple expression

(14) was obtained for the spectral distribution of the signal, which depends on the spectral shape of the laser pulses, quasi-phase mismatches and dispersion characteristics of a crystal. Experimental results, obtained from PPLN crystals with different periods of the nonlinear gratings, are in very good agreement with the theoretical predictions.

This work is supported by a grant from DSTA (Grant No. POD0103451) and EERSS programme, Singapore, and in part by Grant for the leading scientific group of Russia (No. 166.2003.02)

**Captions**

Fig. 1. (a) Transient transmittance changes of probe beam for sample 1 with 30 μm grating period and total sample thickness of 17 mm, gray dash line indicated in inset of (a) is the best-fitting results according to Eq. (13) with pulse width of 200 fs; (b) Power spectrum of Fourier transform of (a). (c) Geometric layout of pump and probe beams relative to the samples.

Fig. 2. (a) Transient transmittance changes of probe beam for sample 2 with 60 μm grating period and total sample thickness of 0.48 mm; (b) Power spectrum of Fourier transform of (a).

Fig. 3. Absorption coefficient $\alpha$ (solid line) and refractive index n (dot line) versus THz frequency $\Omega$ in Mg:$LiNbO_3$ crystal. The curves are calculated with Eq. (15) and (16).

Fig. 4. Calculated THz power spectrum and experimental data for sample 1 (a) and sample 2 (b). The calculation is carried out according to Eq. (14) with pulse duration of 100 (gray dot line), 150 (light gray dash line) and 200 fs (black dash dot line), respectively. The experimental cruve is indicated as dark gray solid line. The dispersion and damping of wave vector in THz region are obtained from Fig. 3.

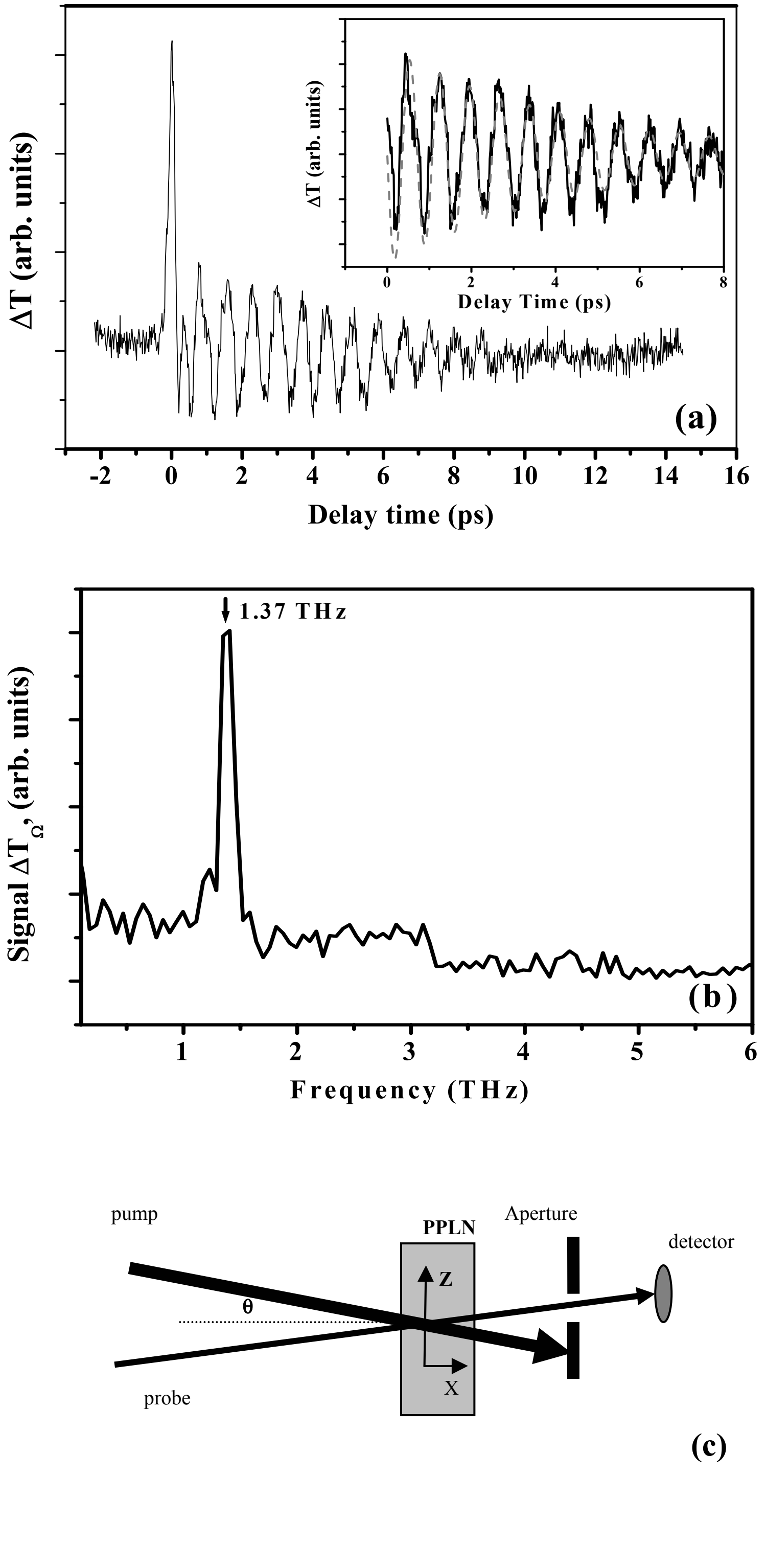

Fig. 1

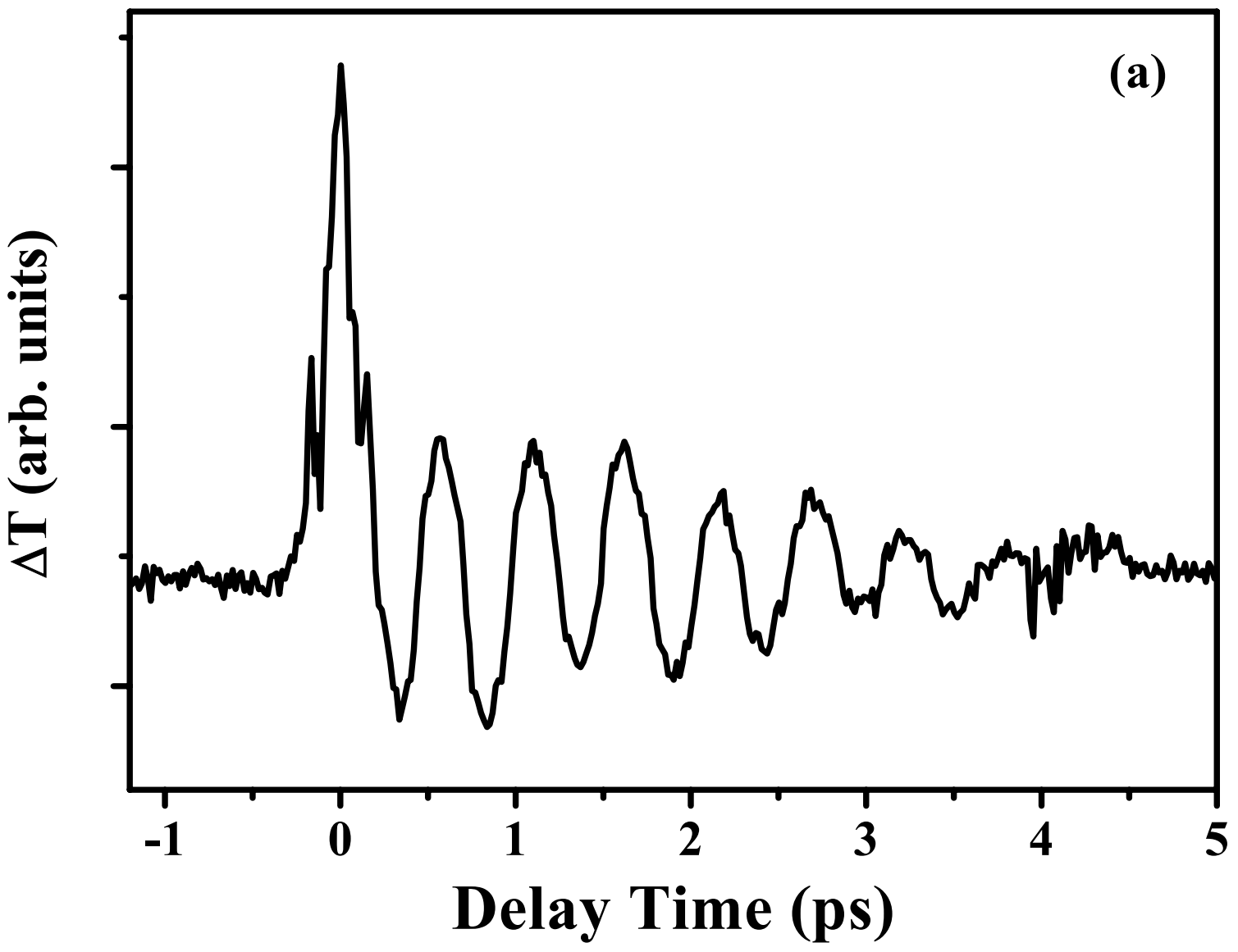

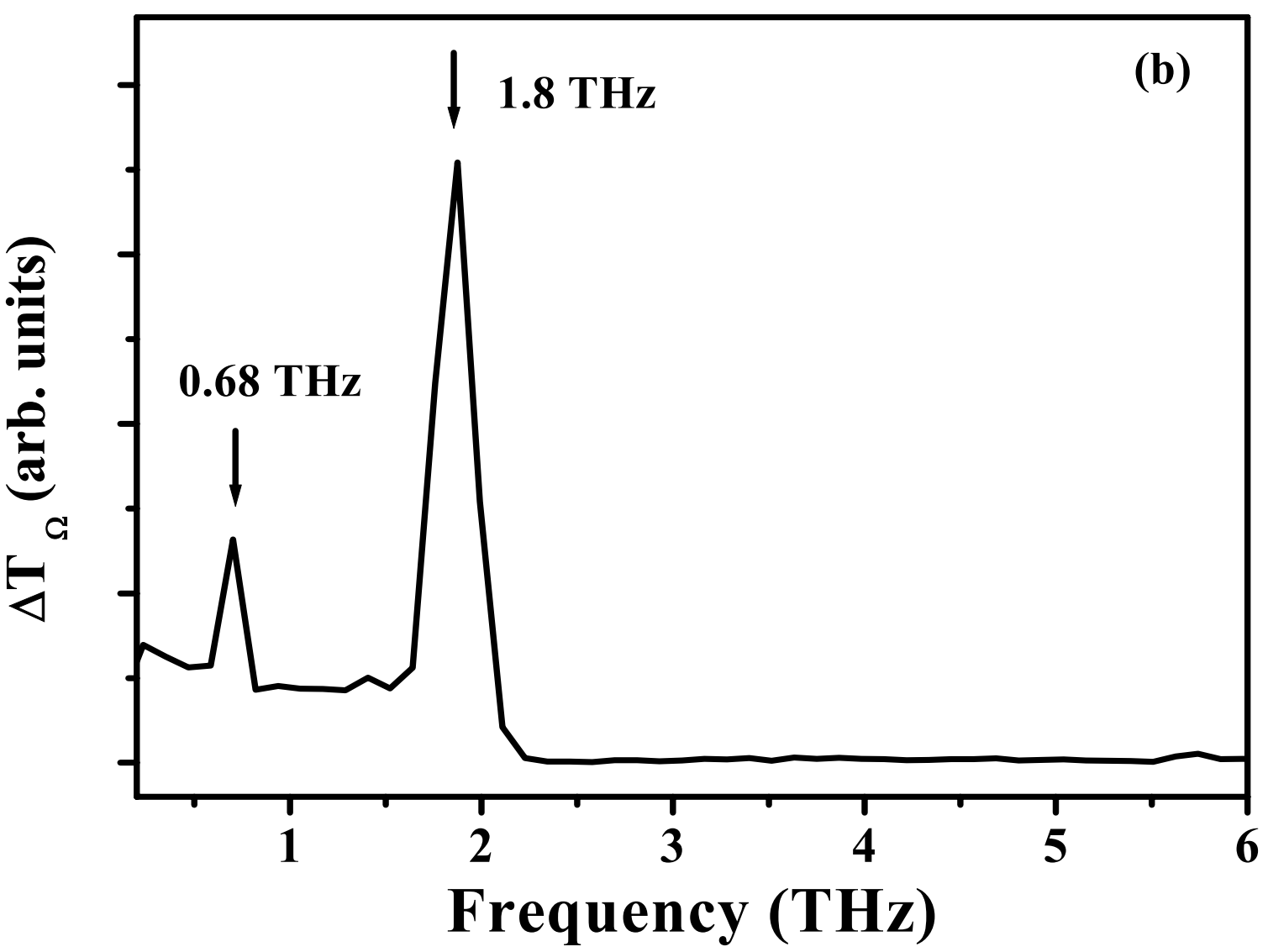

Fig. 2

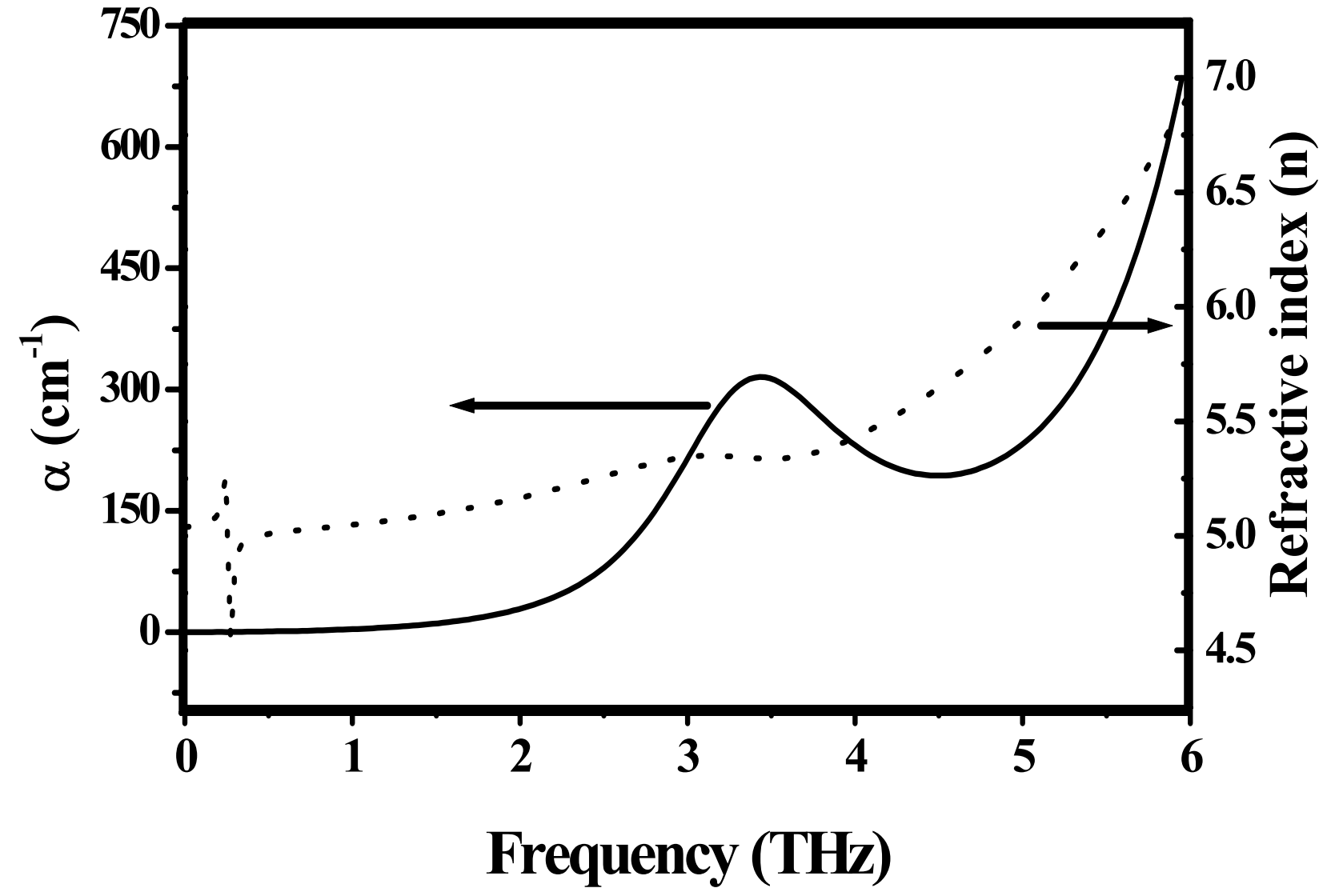

Fig. 3

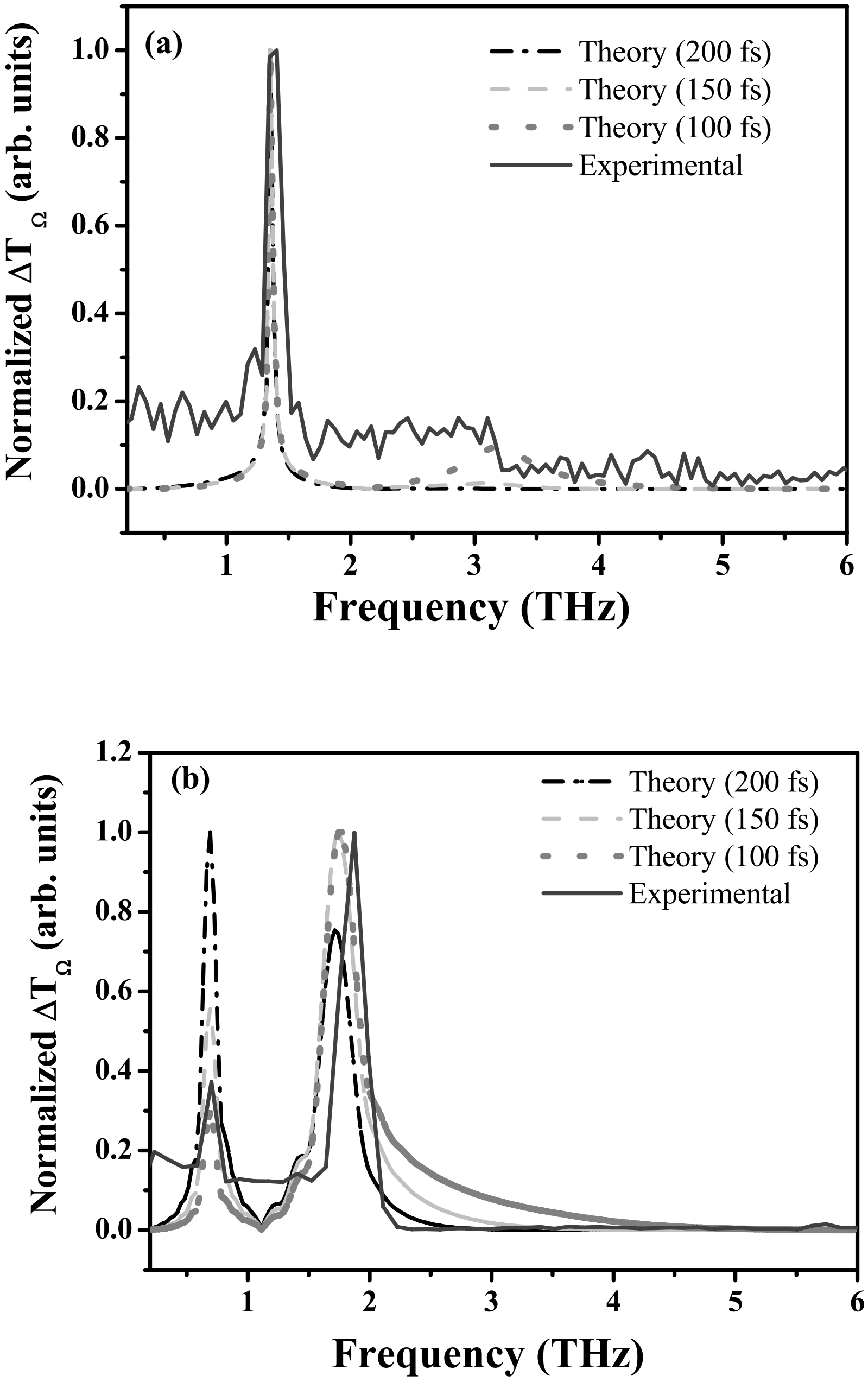

**Fig. 4**